\documentclass[aps,jcp,twocolumn,superscriptaddress,nofootinbib]{revtex4}
\usepackage{amsmath}
\usepackage{amssymb}
\usepackage{textcomp}
\usepackage{graphicx}
\usepackage{dcolumn}
\usepackage{boxedminipage}

\usepackage[usenames]{color}
\definecolor{Red}{rgb}{1.0,0.0,0.0}

\usepackage{verbatim}
\usepackage[colorlinks=true,citecolor=blue,linkcolor=blue]{hyperref}

\graphicspath{{figures/}}

\newcommand{\WW}{\mathbb{W}}
\newcommand{\CC}{\mathcal{C}}

\newcommand{\HH}{\mathbb{H}}

\newcommand{\ee}{\mathrm{e}}

\begin{document}

%%%%%%%%%%%%%%%%%%%%%%%%%%%%%%%%%%%%%%%%%%%%%%%%%%%%%%%%%%%%%
% TITLE
%%%%%%%%%%%%%%%%%%%%%%%%%%%%%%%%%%%%%%%%%%%%%%%%%%%%%%%%%%%%%
\title{Space-time phase transitions in the East model with a softened kinetic constraint}

\author{Yael S. Elmatad}
\thanks{Corresponding author}
\email[Electronic Address: ]{yael@nyu.edu}
\affiliation{Center for Soft Matter Research, Department of Physics, New York University, New York, NY 10003, USA}

\author{Robert L. Jack}
\email[Electronic Address: ]{r.jack@bath.ac.uk}
\affiliation{Department of Physics, University of Bath, Bath BA2 7AY, United Kingdom}

\date{\today}

%%%%%%%%%%%%%%%%%%%%%%%%%%%%%%%%%%%%%%%%%%%%%%%%%%%%%%%%%%%%%
% ABSTRACT
%%%%%%%%%%%%%%%%%%%%%%%%%%%%%%%%%%%%%%%%%%%%%%%%%%%%%%%%%%%%%

\begin{abstract}
The East model has a dynamical phase transition between an active (fluid) and inactive (glass) state.  We show that this phase transition generalizes to ``softened'' systems where constraint violations are allowed with small but finite probabilities.  Moreover, we show that the first order coexistence line separating the active and inactive phases terminates in a finite-temperature space-time critical point.   Implications of these results for equilibrium dynamics are discussed.
\end{abstract}

\maketitle

%%%%%%%%%%%%%%%%%%%%%%%%%%%%%%%%%%%%%%%%%%%%%%%%%%%%%%%%%%%%%
% INTRODUCTION
%%%%%%%%%%%%%%%%%%%%%%%%%%%%%%%%%%%%%%%%%%%%%%%%%%%%%%%%%%%%%
Glasses are formed when liquids are cooled below their melting temperatures.  These supercooled liquids eventually fall out of equilibrium and form glass \cite{Ediger_JPhysChem_1996,Angell_Science_1995}.  The transition between a flowing supercooled liquid and a rigid glass occurs with only a modest change in temperature.  The resulting glass has a microscopic structure which appears to be indistinguishable from its originating supercooled liquid, despite large, but not abrupt, changes in its macroscopic properties.  Several theories connect the glass transition to some kind of unusual equilibrium phase transition \cite{Kirkpatrick1987, Kirkpatrick1987a, Kivelson1995, Gotze1992}.  
More recently, studies based on the dynamical facilitation theory of the glass transition~\cite{Garrahan2003,Chandler2010_Review} have demonstrated connections between glassy behavior and dynamical phase transitions, which take place not in equilibrium ensembles but in space-time ensembles of 
trajectories~\cite{Merolle_PNAS_Aug_2005,Jack2006, Garrahan_PRL_2007, Jack2010,Jack2010b,Hedges_Science_2009, Speck2012}.  Such phase transitions are not typically accompanied by equilibrium phase transitions~\cite{Merolle_PNAS_Aug_2005, Garrahan_PRL_2007}, although some models do support both equilibrium and dynamical transitions~\cite{Jack2010,Jack2010b}. 
The dynamical phase transitions are controlled by a biasing field, 
herein known as $s$, which couples to a measure of space-time activity, $K$.  Here we investigate these phase transitions in
a softened version of the East model~\cite{JJackleandAKronig1994}.  

Glasses are often separated into two categories, labeled fragile and strong \cite{Angell1991}.  Strong glassformers are those whose transport properties (such as relaxation time $\tau$ or viscosity $\eta$) follow an Arrhenius scaling with respect to inverse temperature, $\log \tau \sim  1/T$.   On the other hand, fragile glassformers are those that have transport properties that depend on temperature in a more dramatic ``super-Arrhenius'' fashion.  Within the dynamical facilitation theory~\cite{Garrahan2003,Chandler2010_Review}, strong glass behavior is linked with diffusive relaxation processes while fragile behavior is usually correlated with hierarchical relaxation.  In practice, glasses can exhibit a range of behaviors interpolating between strong and fragile behavior. In fact, it has been shown~\cite{Elmatad2009, Elmatad2010} that relaxation-time data for glassformers can be collapsed onto a single parabolic curve where $ \log \tau \sim 1/T^{2} $. This collapse was theoretically predicted based on kinetically constrained models (KCMs) of glassformers with hierarchical relaxation~\cite{Garrahan2003} suggesting (at least within dynamical
facilitiation theory) that most molecular glassformers exhibit some form of hierarchical relaxation.  The softened East model has such hierarchical relaxation, which motivates its use in this study.  In particular, we will compare the behavior of the softened East model with the softened Fredrickson-Andersen (FA) model considered in~\cite{Elmatad_PNAS_2010}, which has  diffusive relaxation mechanism and scaling reminiscent of strong glasses.  We will discuss similarities between the dynamical phase transitions in these two models, as well as important differences that arise due to the combination of softened kinetic constraints and hierarchical relaxation with super-Arrhenius scaling.

\section{Softened East Model}
\label{sec:model}

The East model is a schematic lattice model for a fragile glassformer~\cite{JJackleandAKronig1994}.  We work in one dimension but we expect our results to generalize trivially to higher dimension.  The thermodynamic properties of the model are those of a lattice gas \cite{IMSM} with $N$ spins that take binary values $n_{i} = \{0,1\}$ and an energy $E = J \sum_{i=1}^N n_i$, where $J$ is the energy required to create an ``excitation'' (a site with $n_i=1$).  The spins do not represent the particles of the glass-former directly: instead they the describe the potential for mobility in a dynamically heterogenous material, after individual particles have been coarse-grained away~\cite{Garrahan2003}.  ``Excitations'' are regions of space where relaxation can (but may not necessarily) occur.  Regions which locally cannot support relaxation are modeled by sites with $n_{i}=0$.  

An excitation on site $i$ relaxes, i.e. $ n_{i} = 1 \rightarrow n_{i} = 0$ with rate $[r_{i}]_{0\rightarrow1} = \lambda C_{i}$, where $C_{i}= \epsilon+n_{i-1}$ is a facilitation function (or kinetic constraint~\cite{Ritort2003}) that depends on the state of the left neighbor site, $n_{i-1}$. 
The facilitation process is such that an excitation on site $i-1$ facilitates motion on its ``eastern'' (right) neighbor, site $i$.
On fixing the excitation energy $J$ and the temperature $T$, the remaining rates in the system are fixed by detailed balance:
$[r_{i}]_{0 \rightarrow 1} = \gamma \lambda C_{i}$ where $\gamma = \ee^{-J/T}$.

In what follows we set $\lambda=1$, which fixes our time unit, and we take Boltzmann's constant $k_\mathrm{B}=1$ throughout.
Having fixed our units in this way, the system at equilibrium is described by two dimensionless parameters, $\gamma$ and $\epsilon$.  
We imagine that violating the kinetic constraint requires an energy $U$, so that $\epsilon = \ee^{-U/T}$.  Hence, one may equivalently consider the system as a function of the reduced temperature $T/J$ and the ratio of energies $U/J$.  

In terms of the equilibrium dynamics on decreasing the temperature (at fixed $U$ and $J$), the soft East model exhibits two crossovers.  
At high temperature, facilitation effects are irrelevant, as usual in KCMs~\cite{Ritort2003}, with facilitation becoming relevant around the onset temperature
$T_\mathrm{o} \approx J$.  However, in the presence of a soft constraint, facilitation effects are also irrelevant at very low temperatures, as we now discuss. 
The facilitated dynamics of the ``hard'' East model (with $\epsilon=0$) lead to a time scale $\tau_{\rm fac}(\ell)$ for dynamical relaxation on a length scale $\ell$: the relevant rate is 
\begin{equation}
\tau_{\rm fac}(\ell)^{-1} \sim \exp( -(J/T) \log_2 \ell),
\end{equation}
which is valid as long as $\ell < \ee^{J/T}$ and $J/T\ll 1$.
The equilibrium relaxation time of the model with $\epsilon=0$ scales as  $\tau_\mathrm{East} \sim \ee^{J^2/(2 T^2\ln 2)}$~\cite{Cancrini2007}, comparable with the prediction~\cite{Sollich1999} obtained by setting $\ell$ equal to the typical spacing between excitations, $ \ee^{J/T}$.  When $\epsilon>0$,
one expects an additional relaxation mechanism via constraint violation, with rate $\tau_{\rm soft}^{-1} \sim \epsilon \sim \exp (-U/T)$.  
Thus, the structural relaxation time $\tau$ in the soft East model may be estimated via  $\tau^{-1} \approx \tau_{\rm East}^{-1} + \tau_{\rm soft}^{-1}$. 

At low temperatures, the facilitated time scale diverges faster than $\tau_\mathrm{soft}$, and the softened mechanism dominates. (This is opposite to the behavior 
in the soft-FA model, where the facilitated relaxation dominates as $T\to0$, as long as $J>3U$~\cite{Elmatad_PNAS_2010}.)
The crossover to the softened mechanism cuts off the super-Arrhenius divergence of $\tau$ in favor of an Arrhenius law, reminiscent of strong glasses.  This crossover takes place at 
\begin{equation}
T_\mathrm{x} \approx \frac{J^2}{2 U \ln 2}
\end{equation}
and is accompanied by a saturation of the equilibrium dynamical length scale at $\xi_\mathrm{sat} \approx 2^{U/J}$.  However, in contrast to the crossovers from fragile to strong behavior in other KCMs~\cite{Buhot2001,Garrahan2003}), we note that the activation energy for relaxation here increases monotonically with decreasing temperature.  In experiments on glass-formers, we expect $U$ to be a large energy scale, so that $T_\mathrm{x}$ is likely to be beyond the limit of supercooling of the liquid.  However, this low temperature crossover will be relevant for our discussion of phase transitions in the soft-East model below.

%%%%%%%%%%%%%%%%%%%%%%%%%%%%%%%%%%%%%%%%%%%%%%%%%%%%%%%%%%%%%
% Active and Inactive Space-Time Phases
%%%%%%%%%%%%%%%%%%%%%%%%%%%%%%%%%%%%%%%%%%%%%%%%%%%%%%%%%%%%%

\section{Active and Inactive Space-Time Phases}

In order to investigate dynamical (space-time) phase transitions in the soft East model, we define an order parameter $K$ that measures the activity in the system~\cite{Merolle_PNAS_Aug_2005, Garrahan_PRL_2007, Lecomte2005,Lecomte2007,Hooyberghs2010, Baiesi2009}.
The activity is the total number of configuration changes in a trajectory,
$K = \sum_{i=1}^N \sum_{t=0}^{t_\mathrm{obs}} \left (n_i(t) - n_i(t+\delta t) \right)^2$, where $t_{\mathrm{obs}}$ is the length (``observation time'') of the trajectory, and $\delta t$
is an elementary time step.
Since we work in continuous time \cite{Newman1999} we take the limit where $\delta t \rightarrow 0$.

\begin{figure}[b]
\resizebox{8cm}{!}{\includegraphics{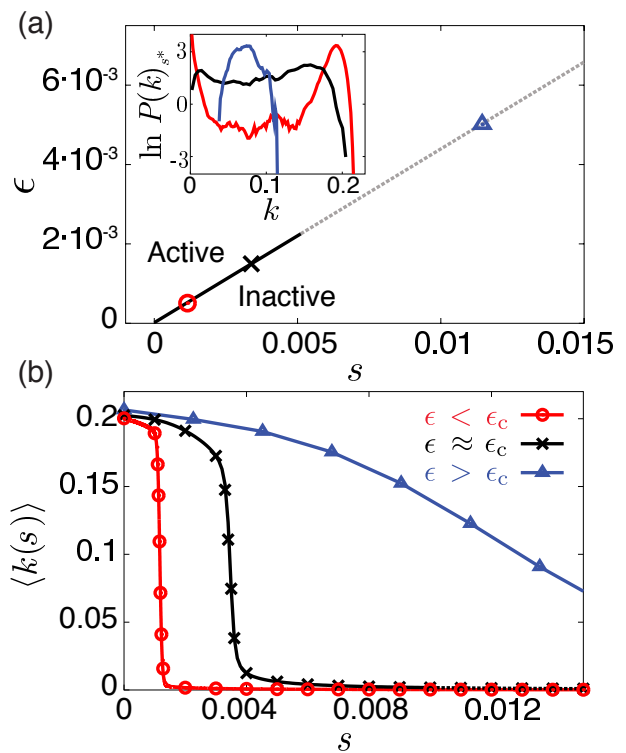}}
%{!}{\includegraphics{fig3.pdf}}
\caption{
(A) Phase diagram for the 1$d$ soft East model in the $(s, \epsilon)$ plane at $J/T=0.75$ (so $\gamma = 0.47$).  The solid line is the phase boundary between the active and inactive phase.  The dashed line is the continuation of the symmetry line  (\ref{eqn:sym}) into the 1 phase region. The red circles indicate a state point with $\epsilon = 5 \cdot 10^{-4} < \epsilon_\mathrm{c}$, on the coexistence line, in the two phase region.  The blue triangles indicates a simulation point where $\epsilon = 5 \cdot 10^{-3} > \epsilon_\mathrm{c}$.  The black $\times$  indicates a state near the critical point: $\epsilon = 1.5 \cdot 10^{-3} \approx \epsilon_\mathrm{c}$. (The precise location of the critical point is not known for this model).  The inset to (A) shows  histograms of the intensive activity $k$ for the three state points in (A). (B) Plots of average intensive activity $\langle k(s) \rangle$ as a function of field $s$, for the values of $\epsilon$ given in (A). 
\label{fig:phaseD}}
\end{figure}

\begin{figure*}[tb]
\resizebox{6.5in}{!}{\includegraphics{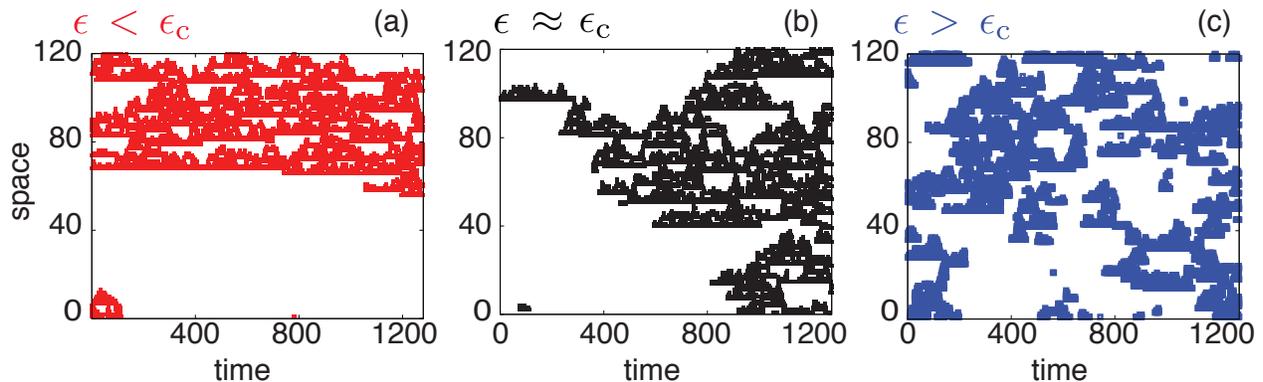}}
\caption{Sample trajectories from the three state points identified in Fig. \ref{fig:phaseD}, taken from near the centers of the distributions $P(k)$ ($k\approx 0.1$).  Thus, for $\epsilon \leq \epsilon_\mathrm{c}$ these trajectories are rare, coming from the trough in the histogram that lies between the two stable basins. Active sites are colored ($n_i = 1$) and inactive sites are white ($n_i=0$).  (A) Trajectory with $\epsilon < \epsilon_\mathrm{c}$. showing space-time phase separation.  (B) Trajectory at $\epsilon \approx \epsilon_\mathrm{c}$ where the phases are still identifiable but the clusters no longer have a sharp interface.  (C) Trajectory at $\epsilon > \epsilon_\mathrm{c}$ showing a single homogeneous phase.
\label{fig:trajs}}
\end{figure*}

The activity $K$ distinguishes between active and inactive phases  (the active phase is characterized by a large value of $K$ while the inactive phase is characterized by a small value of $K$).  While $K$ is an order parameter in space time, it is analogous to an equilibrium order parameter such as density, which distinguishes between two phases such as liquid and gas.  However, while the equilibrium ensemble for a liquid-to-gas phase transition is a set of configurations, the ensembles we consider here are collections of space-time trajectories.  (Mathematically, we are considering the ``large deviations'' of the order parameter $K$~\cite{Lecomte2007,Garrahan_JPhysA_2008}.)

To access space-time phase transitions, we introduce an intensive biasing field $s$ which couples to $K$.  This defines a nonequilibrium ensemble of trajectories \cite{Ruelle1984, Merolle_PNAS_Aug_2005, Jack2006, Garrahan_PRL_2007, Lecomte2005, Lecomte2007} known as the $s$-ensemble \cite{Garrahan_JPhysA_2008}. For an observable $A$, we use $\langle A \rangle_{s}$ to denote the expectation value of  $A$ in the presence of a biasing field of strength $s$.  Similarly, $\langle A \rangle_{0}$ denotes the equilibrium expectation value.  These expectation values are related through
\begin{equation}
\langle A \rangle_{s} =\langle A e^{-sK} \rangle_{0} \frac{1}{Z(s,t_\mathrm{obs})} 
\label{eqn:avgAs}
\end{equation}
where $Z(s,t_{\mathrm{obs}}) = \langle \exp(-sK) \rangle_{0}$ is the partition function for the $s$ ensemble.

While the field $s$ has no direct physical interpretation, the formalism we use implies an equivalence between ensembles with fixed $K$ and those with fixed $s$~\cite{Garrahan_JPhysA_2008}.  That is, the field $s$ acts in the same way as a constraint on $K$.  

%%%%%%%%%%%%%%%%%%%%%%%%%%%%%%%%%%%%%%%%%%%%%%%%%%%%%%%%%%%%%
% Computational Sampling of Space-Time Phases
%%%%%%%%%%%%%%%%%%%%%%%%%%%%%%%%%%%%%%%%%%%%%%%%%%%%%%%%%%%%%

\section{Computational Sampling of Space-Time Phases}

To harvest ensembles of trajectories we use transition path sampling (TPS) \cite{Bolhuis_AnnuRevPhysChem_2002}, as in~\cite{Hedges_Science_2009,Elmatad_PNAS_2010}.  
We employ two basic move types to generate trajectories from an initial trajectory: shooting and shifting in both the forwards and backwards directions.  New trajectories are accepted with a probability proportional to $\exp(-sK)$ - analogous to accepting and rejecting configurations with probability proportional to $\exp(-\beta E)$ in standard configurational Monte Carlo dynamics.

For the softened East model, the $s$-ensemble depends on only three parameters: $\gamma, \epsilon$, and $s$.  Fig.~\ref{fig:phaseD}(a) shows a space-time phase diagram in the $(\epsilon,s$) plane, for fixed $\gamma$ (specifically, $J/T=0.75$, so $\gamma= 0.47$).  On the $s=0$ axis, the system undergoes equilibrium dynamics, while the $\epsilon=0$ axis is the `hard' East model.
The solid line on the phase digram indicates
phase coexistence between inactive and active phases, as estimated from our numerical results.  On increasing $s$, one crosses the phase coexistence line at $s=s^*$ and the activity in the system changes discontinuously from an ``active'' state to an inactive one: this is accompanied by a jump in the value of $K$ [see Fig.~\ref{fig:phaseD}(b)].

To characterize these phase transitions in the $s$-ensemble, we perform finite-size scaling, by varying the system size $N$ and the observation time $t_\mathrm{obs}$.  As in~\cite{Elmatad_PNAS_2010}, there are two analytic results that improve the quality of our finite-size scaling analysis.  The theoretical analysis is given in Sec.~\ref{sec:theory}: here we simply state the results and explain how we use them in conjunction with our TPS scheme. Firstly,
we perform our TPS simulations at $s=s^{*}$, which allows accurate estimation of the properties of the phase transition.   The transition point $s^*$ can be located exactly because
there is a hidden symmetry of the $s$-ensemble for the soft East model, similar to that found in the soft-FA case~\cite{Elmatad_PNAS_2010}. This implies that $s^*$ satisfies
\begin{equation}
\frac{1+\gamma}{2\epsilon+1} = \sqrt{(1-\gamma)^{2} + 4\gamma e^{-2s^{*}}} \ \ \ .
\label{eqn:sym}
\end{equation}

A second result that aids finite-size scaling analysis involves the boundary conditions used in the $s$-ensemble: as in~\cite{Elmatad_PNAS_2010},  it is natural to take periodic boundary conditions for the $N$ spins in the system, but the boundary conditions in time are not periodic, with the initial and final configurations of the trajectory being free to fluctuate independently. This free boundary favors activity near initial and final parts of the trajectory~\cite{Garrahan_JPhysA_2008}, which frustrates convergence of the limit of large $t_\mathrm{obs}$.  To counteract this effect, we bias the initial and final conditions of the trajectory, as follows.  We define an angle $\alpha$ by
\begin{equation}
\tan \alpha = \frac{2e^{-s}\sqrt{\gamma}}{1-\gamma}
\label{equ:alpha}
\end{equation}
with $0<\alpha<\pi/2$,
and we define
\begin{equation}
g_\mathrm{sEast} = \ln\left[ \tan (\alpha^*/4) \sqrt{\gamma} \right] \ \ \ .
\label{eqn:softEastBoundary}
\end{equation}
where $\alpha^*$ is the value of $\alpha$ when $s=s^*$.
We then introduce an extra bias on the $s$-ensemble, which depends on the total number of excitations in the initial and final conditions
of the trajectory: in the resulting ensemble then the average of observable $A$ is
\begin{equation}
\langle A \rangle_{s,\text{symm}} = \frac{\langle A e^{-sK + g_\mathrm{sEast}[\mathcal{N}(0)+\mathcal{N}(t_{\mathrm{obs}})]}\rangle_{0}}{Z_{\text{sym}}(s,t_{\mathrm{obs}})},
\label{equ:s-sym}
\end{equation}
where $\mathcal{N}(t)  = \sum_{i=1}^N n_i (t)$ is the number of excitations at time $t$, and
$Z_{\text{sym}}(s,t_{\mathrm{obs}}) = \langle e^{-sK + g_\mathrm{sEast}[\mathcal{N}(0)+\mathcal{N}(t_{\mathrm{obs}})]}\rangle_{0}$ is a normalization
factor.
As discussed in~\cite{Elmatad_PNAS_2010},  this extra bias ensures that the $s$-ensemble at $s=s^*$ is fully symmetric between the active and inactive phases, and we term it the symmetrized $s$-ensemble.

In the limit where the observation time $t_\mathrm{obs}$ is infinite, this symmetrized ensemble is equivalent to the ensemble without boundary constraints, 
$\langle A \rangle_{s,\text{symm}} \to \langle A\rangle_s$ as $t_\mathrm{obs}\to\infty$~\cite{Elmatad_PNAS_2010}, as long as observable $A$ is not dominated by the initial and final parts of the trajectory.  
This boundary biasing condition is straightforwardly included in the transition path sampling algorithm.

\section{Numerical results}

The order parameter for the phase transitions we consider is the intensive average activity $\langle k (s) \rangle$,
\begin{equation}
\langle k(s) \rangle \equiv \frac{1}{N t_{\mathrm{obs}}} \langle K \rangle_{s,\mathrm{symm}} \ \ \ .
\end{equation}
As discussed in the previous Section,
our main results are summarized in Fig.~\ref{fig:phaseD} where
we show sample numerical results for $k(s)$ (panel b), and the phase diagram (panel a) that we have obtained by finite-size scaling of the behavior of $k(s)$.  In particular, there is a first-order space-time phase transition represented by a solid line, which ends in a critical point as $(s_c,\epsilon_c)$.  That is, the phase transition that is known to be present as $\epsilon=0$~\cite{Garrahan_PRL_2007} still exists in the presence of soft constraints, as long as $\epsilon < \epsilon_c$.  However, when the constraints are too soft ($\epsilon>\epsilon_c)$, the phase transition disappears and the system shows a smooth response to the field $s$, and no phase transition.  The exact value of $\epsilon_c$ is not known for the soft East model, but we have bracketed its location by identifying two state points, one where the phase transition occurs, and another where the response to $s$ is smooth (non-singular).
 The inset to Fig. \ref{fig:phaseD}(a) shows probability distributions $P(k)$, where $k=K/Nt_{\mathrm{obs}}$ is the (intensive) activity per space-time volume.  The histograms correspond to simulations performed at the conditions highlighted by the symbols along the symmetry line.  These were chosen to lie in the two phase region far from criticality, near criticality, and in the one phase region which does not support a phase transition.   In the first order region $\epsilon < \epsilon_\mathrm{c}$, there are two distinct peaks corresponding to the inactive and active phases.  Near the critical point $\epsilon \approx \epsilon_\mathrm{c}$ the two peaks are broadened and very flat.  In the one phase region $\epsilon > \epsilon_\mathrm{c}$ there is one distinct peak.  

Figure \ref{fig:phaseD}(b) shows the jump in $\langle k(s) \rangle$ as the field $s$ is varied.  It is useful to compare these results with those expected for an equilibrium phase transition in a  ferromagnetic system (see also~\cite{Merolle_PNAS_Aug_2005,Jack2006}).
 The behavior of $k(s)$ is analogous to the change in the magnetization, $M$, of a ferromagnet as the field strength, $h$, is changed.
 Thus, $\epsilon < \epsilon_\mathrm{c}$ corresponds to $T<T_\mathrm{c}$ in a ferromagnet and shows a jump in the order parameter (from a high $k$ active phase to a low $k$ inactive phase).  On the other hand, $ \epsilon \approx \epsilon_\mathrm{c}$ corresponds to $T \approx T_\mathrm{c}$ in a ferromagnet: it  shows a steep crossover between the phases (for the system sizes considered here). 

In Fig.~\ref{fig:trajs}, we show three trajectories which have $k \approx 0.1$, approximately corresponding with the center of the distributions. These trajectories are harvested under the same conditions as the marked points in Fig. \ref{fig:phaseD}(a). 
In Fig. \ref{fig:trajs}(a) a trajectory for $\epsilon < \epsilon_\mathrm{c}$ is shown: it may be seen from the inset to Fig \ref{fig:phaseD}(b) that
this is a rare trajectory, coming from near the minimum of $P(k)$.  There is a clear segregation between the two phases as the space-time ``surface tension" \cite{Elmatad_PNAS_2010} is large.  In Fig. \ref{fig:trajs}(b) a near-critical trajectory with $\epsilon \approx \epsilon_\mathrm{c}$ is shown.  Here, there is still distinct phase segregation but the boundaries between the phases have become amorphous, corresponding to the diminished ``surface tension'' near criticality.  In Fig. \ref{fig:trajs}(c) we show a trajectory with $\epsilon > \epsilon_\mathrm{c}$. Here, there is no phase separation since the system
in a one-phase region of the phase diagram.

\begin{figure}[b]
\resizebox{8cm}{!}{\includegraphics{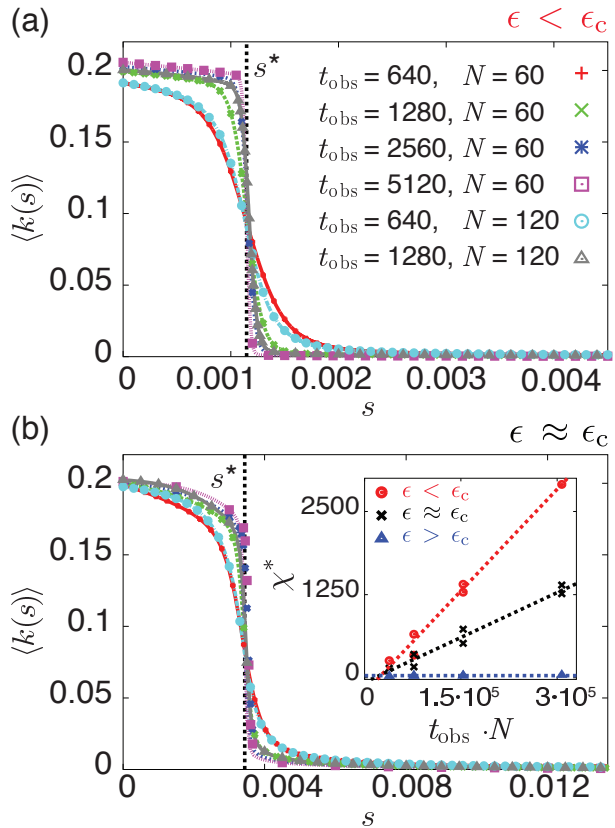}}
\caption{
Finite size scaling of the intensive activity $\langle k(s) \rangle$. In panel (a) we show  $\epsilon =  5 \cdot 10^{-4} < \epsilon_\mathrm{c}$ and in (b) $\epsilon = 1.5 \cdot 10^{-3} \approx \epsilon_\mathrm{c}$. Inset to (B) shows 
scaling of the susceptibility $\chi^* = -d \langle k \rangle/ds \big{|}_{s^{*}} $ for various system sizes $t_\mathrm{obs} \cdot N$, with symbols and colors corresponding to the$\epsilon$ values  shown in Fig.~\ref{fig:phaseD}.  The dashed lines indicates linear scaling associated with a first order phase transition (for $\epsilon\leq\epsilon _\mathrm{c}$ and the constant value of susceptibility expected in a single phase ($\epsilon > \epsilon_\mathrm{c}$). 
\label{fig:finSize}}
\end{figure}

As usual for phase transitions in space-time~\cite{Garrahan_JPhysA_2008}, the first order jumps in $k(s)$ become singular only as $N,t_\mathrm{obs} \rightarrow \infty$.   For finite systems undergoing a first-order phase transition, the susceptibility $\chi = -d \langle k (s) \rangle/ds$ evaluated at $s^*$ is proportional to the system size, $N t_\mathrm{obs}$ \cite{Binder1984}.  This scaling is shown in Fig. \ref{fig:finSize}.  In Fig. \ref{fig:finSize}(a) we show the finite size scaling for the first order transition for $\epsilon < \epsilon_\mathrm{c}$.  As the system size increases, the transition sharpens.  In Fig. \ref{fig:finSize}(b) we show the scaling near the critical point.  Here, the crossover sharpens as $N t_\mathrm{obs}$ increases, but there is no clear jump in $\langle k(s) \rangle$.  In the inset to Fig. \ref{fig:finSize}(b) we plot $\chi^*$, the value of the susceptibility evaluated at the transition point $s^*$ as function of system size $N t_\mathrm{obs}$.   For $\epsilon < \epsilon_\mathrm{c}$ and $\epsilon \approx \epsilon_\mathrm{c}$ the scaling grows linearly.  For $\epsilon > \epsilon^*$ there is no increase with system size. 
%%%%%%%%%%%%%%%%%%%%%%%%%%%%%%%%%%%%%%%%%%%%%%%%%%%%%%%%%%%%%
% Theoretical Analysis
%%%%%%%%%%%%%%%%%%%%%%%%%%%%%%%%%%%%%%%%%%%%%%%%%%%%%%%%%%%%%

\section{Theoretical Analysis}
\label{sec:theory}

\subsection{The master equation and its symmetries}

The soft East model may be studied analytically following the methods used for the soft FA model in~\cite{Elmatad_PNAS_2010}.
We now use these methods to derive the relations (\ref{eqn:sym}) and (\ref{eqn:softEastBoundary}) that we used to facilitate our numerical studies. 
To do so, we start with the master equation
\begin{equation}
\partial_t P(\CC,t) = -r(\CC) P(\CC,t) + \sum_{\CC'(\neq \CC)} W (\CC' \rightarrow \CC) P(\CC',t) \ \ \ 
\label{eqn:masterEast1}
\end{equation}
where $P(\CC,t)$ is the probability of observing a configuration $\CC$ of the soft east system at time $t$. Also, $r(\CC) = \sum_{\CC'} W(\CC \rightarrow \CC')$ is the exit rate to leave the current configuration $\CC$, sums over $\CC'$ run over all possible configurations of the model, and $W(\CC' \rightarrow \CC)$ is the transition rate from $\CC'$ to $\CC$.

This master equation can be represented compactly using a spin-half representation of the system's configuration space.  The ground state (where $n_i = 0$ for all $i$) is denoted  by $| \Omega \rangle$, with the site variables $\{ n_i \}$ now represented by $N$ spin-half variables.  Hence, all possible configurations $\{ n_i \}$ can be represented via Pauli matrices $\sigma_i^{x,y,z}$ acting on $|\Omega\rangle$ \cite{Sakurai1993}, the raising $\sigma_i^+$ and lowering $\sigma_i^-$ operators are defined as $\sigma_i^\pm = \frac{1}{2} (\sigma_i^x \pm \sigma^y_i$)
Thus, a configuration $\{n_i\}$ can be represented as
\begin{equation}
| \{ n_i\} \rangle = \sum_{i=1}^N (\sigma_i^{+})^{n_i} | \Omega \rangle \ \ \ .
\end{equation}
We then write
$
| P(t) \rangle = \sum_{\CC} P(\CC,t) | \CC \rangle
$
which allows us to rewrite the master equation (\ref{eqn:masterEast1}) as
\begin{equation}
\frac{\partial}{\partial t} | P(t) \rangle = \WW | P(t) \rangle \ \ \ ,
\end{equation}
where $\WW$ is a linear operator whose matrix elements are the transition rates, $W(\CC\to\CC')$ and escape rates $r(\CC)$.

For the soft East model this operator is 
\begin{equation}
\WW = \sum_{i}(\hat{n}_{i-1} + \epsilon) [(1-\sigma_i^+)\sigma_i^- + \gamma (1-\sigma_i^-)\sigma_i^+]  \ \ \ .
\label{eqn:WWs0East}
\end{equation}
This master operator is similar to that of the soft-FA model given in Ref. \cite{Elmatad_PNAS_2010}. However it does not contain the symmetrized portion ( $i \leftrightarrow j$ ) as the East model is inherently asymmetric, nor does it contain the diffusive term that was introduced to the soft-FA model: introducing such a term in the soft-East model does not simplify the analysis in the way that it did for the soft-FA model so we do not consider it here.

To analyse the soft East model in the $s$-ensemble, we follow \cite{Garrahan_JPhysA_2008}.  
First the probability distribution $P(\CC,K,t)$ is defined as the probability of being in configuration $\CC$ at time $t$, having already accumulated an activity $K$ in the time between time $0$ and $t$.   The probability to be in configuration $\CC$ at time $t$ at a field strength $s$ is
then $P(\CC,s,t)$, which can then be written as a reweighted sum over over all possible accumulated $K$ values:
\begin{equation}
P(\CC,s,t) = \sum_K P(\CC,K,t)e^{-sK} \ \ \ .
\end{equation}
The equation of motion for $ | P(s,t) \rangle = \sum_{\CC} P(\CC,s,t) | \CC \rangle $ is
\begin{equation}
\frac{\partial}{\partial t} | P(s,t) \rangle = \WW(s) | P(s,t) \rangle \ \ \ ,
\end{equation}
where now the matrix elements of $\WW(s)$ include the biasing field $s$, through 
\begin{equation}
\WW(s) = \sum_i  (\hat{n}_{i-1} + \epsilon) [(e^{-s}-\sigma_i^+)\sigma_i^- + \gamma(e^{-s}-\sigma_i^-)\sigma_i^+]  \ \ \ .
\end{equation}
This resembles Eq.~(\ref{eqn:WWs0East}), with the additional factors of $e^{-s}$ modifying the rate of changes of state.  

This representation of the model allows us to derive the symmetry condition (\ref{eqn:sym}), 
following \cite{Jack2006b,Elmatad_PNAS_2010}. Writing the energy operator $\mathbb{E} = J\sum_i \hat{n_i}$ 
(recall $\gamma = \ee^{-J/T}$), we symmetrize $\WW(s)$ using $\HH(s) \equiv \ee^{\mathbb{E}/2T} \WW(s)  \ee^{-\mathbb{E}/2T}$,
 so that
\begin{multline}
\HH(s) 
= - \frac{h}{4} \sum_i \left( 1 + 2\epsilon + \sigma_{i-1}^z \right) \\  \times
  \left( \frac{1+\gamma}{h} + \sigma_{i}^z \cos\alpha -  \sigma_{i}^x \sin\alpha \right)
  \label{equ:Heast}
\end{multline}
where $h=\sqrt{(1+\gamma)^2 - 4\gamma(1-\ee^{-2s})}$ and $\alpha$ was defined in (\ref{equ:alpha}): we have $(\cos\alpha,\sin\alpha) = \frac{1}{h}( 1-\gamma,2\ee^{-s}\sqrt{\gamma})$.
Next, making a (site-independent) rotation of the quantum spin matrices, 
$(\sigma^x,\sigma^y,\sigma^z) \to ( -\sigma^x\cos\alpha - \sigma^z\sin\alpha , -\sigma^y, \sigma^z\cos\alpha -  \sigma^x\sin\alpha )$, one finds
$\HH(s) \to \HH_\mathrm{W}(s)$ with
\begin{multline}
 \HH_\mathrm{W}(s) = - \frac{h}{4} \sum_i \left( 1 + 2\epsilon + \sigma_{i-1}^z \cos\alpha -  \sigma_{i-1}^x \sin\alpha \right) 
 \\ \times
  \left( \frac{1+\gamma}{h} +\sigma_{i}^z  \right)
  \label{equ:Hwest}
\end{multline}
Comparison with (\ref{equ:Heast}) shows that (\ref{equ:Hwest}) is a symmetrized master operator for a ``West model'', similar to the East model except
that spin $i$ facilitates motion on its left (west) neighbor, spin $i-1$.  This West model has the same parameters as the original East model if
$1 + 2\epsilon = \frac{1+\gamma}{h}$: this corresponds to the symmetry condition (\ref{eqn:sym}).  The spin rotation and the symmetrization of $\WW(s)$ preserve
the eigenspectrum of this operator, meaning that the model is `self-dual', with a 
symmetry that may be spontaneously broken.  Formally, there is a similarity transformation 
\begin{equation}
\HH_\mathrm{W}(s) = U_\alpha^\dag \HH(s) U_\alpha
\label{equ:simil}
\end{equation}
where $U_\alpha=\prod_j (-\ee^{i\alpha\sigma_j^y/2}\sigma^z_j )$: see also~\cite{Elmatad_PNAS_2010,Jack2006b}. When the West model described by $\HH_\mathrm{W}(s)$ has the same parameters as the East model described by $\HH(s)$, this represents a symmetry of the model, which is broken spontaneously at the spacetime phase transitions.

To understand the boundary bias introduced in (\ref{equ:s-sym}), it is useful to write $Z(s,t_\mathrm{obs}) = \langle e | e^{\WW(s) t_\mathrm{obs}} |\mathrm{eq}\rangle$, with $\langle e | = \langle 0 | \prod_j ( 1 + \sigma^-_j)$ and $|\mathrm{eq}\rangle = \prod_j \frac{ 1 + \gamma \sigma^+_j}{1+\gamma} | \Omega \rangle$.  We also introduce the general coherent state $|\phi\rangle = \prod_j ( \cos\phi + \sigma^+_j \sin\phi ) |\Omega\rangle$, parameterized by the angle $\phi$.  Then, symmetrizing $\WW$, one arrives at
\begin{equation}
Z(s,t_\mathrm{obs}) = \langle \phi_0 | e^{\HH(s) t_\mathrm{obs}} | \phi_0 \rangle
\label{equ:Zphi}
\end{equation}
with $\tan\phi_0 = \sqrt{\gamma}$ and $0<\phi_0<\pi/2$.  From (\ref{equ:simil}), one has
\begin{equation}
Z(s,t_\mathrm{obs}) = \langle (\alpha/2) - \phi_0 | e^{\HH_\mathrm{W}(s) t_\mathrm{obs}} | (\alpha/2) - \phi_0 \rangle
\label{equ:Zphi2}
\end{equation}
where we used $U_\alpha^\dag|\phi\rangle = |(\alpha/2)-\phi\rangle$, which follows from the definitions of $U_\alpha$ and $|\phi\rangle$.

The key point is that the partition function (\ref{equ:Zphi}) is fully symmetric under this similarity transformation only if two conditions are met:
(i)~$\HH_\mathrm{W}(s)$ should describe a West model with the same parameters as $\HH(s)$ and (ii)~the boundary terms are invariant
under the transformation: $(\alpha/2) - \phi_0 = \phi_0$, so that (\ref{equ:Zphi2}) coincides with (\ref{equ:Zphi}), up to the replacement
of $\HH_\mathrm{W}(s)$ by $\HH_(s)$.  It is easily verified that while (\ref{eqn:sym}) ensures condition (i),
it does not simultaneously ensure condition (ii).  To simultaneously satisfy both conditions, we write
\begin{equation}
Z_\mathrm{sym}(s,t_\mathrm{obs}) = \langle e | e^{g_\mathrm{sEast}\sum_i \hat{n}_i} e^{\WW(s) t_\mathrm{obs}} e^{g_\mathrm{sEast}\sum_i \hat{n}_i} |\mathrm{eq} \rangle
\end{equation}
Symmetrizing $\WW(s)$ and using the expression (\ref{eqn:softEastBoundary}) for $g_\mathrm{sEast}$, one finds
\begin{equation}
Z_\mathrm{sym}(s,t_\mathrm{obs}) = y^{-N} \langle \alpha/4 |  e^{\HH(s) t_\mathrm{obs}}  |\alpha/4 \rangle 
\end{equation}
with $y = (1+\gamma)\cos^2(\alpha/4)$.  Applying (\ref{equ:simil}) yields $Z_\mathrm{sym}(s,t_\mathrm{obs}) = c^{-N} \langle \alpha/4 |  e^{\HH_\mathrm{W}(s) t_\mathrm{obs}}  |\alpha/4 \rangle$, indicating that the partition function is fully symmetric under the similarity transformation, as long as (\ref{eqn:sym}) holds.  Physically, this means that the initial and final conditions in the symmetrized $s$-ensemble have no additional preference for the active phase over the inactive one, ensuring that the probabilities of active and inactive states are equal within the ensemble that we sample.  This facilitates accurate numerical characterization of the phase transitions in the system.

\subsection{Scaling of $s^*$ and $\epsilon_c$ with temperature}

As well as exact symmetries of the master operator, we also identify low-temperature scaling properties of the softened East model.
For low temperatures, $\gamma \ll 1$, the symmetry condition (\ref{eqn:sym}) reduces to $s^* \approx 2\epsilon/\gamma$.  As proposed in~\cite{Jack-melt-2011}, this scaling of $s^*$ with system parameters may be obtained by comparing the difference $\Delta k$ in activity between the two phases, and the escape rate $\gamma_0$ from the inactive phase.  Specifically, we expect
\begin{equation}
s^* \approx \gamma_0 / \Delta k.
\label{equ:s-soft}
\end{equation}
At low temperatures, the activity difference between the phases is given simply by the equilibrium activity per site, so $\Delta k = 2\gamma^2\lambda$.  The escape rate (per site) is $\gamma_0 \approx \epsilon\lambda\gamma$,  since a single unfacilitated spin flip in a large inactive region typically leads to relaxation of the entire inactive region.  Thus, the prediction
(\ref{equ:s-soft}) works well in the softened East model.  (The same analysis also applies in the softened FA model, except that $\Delta k$ is larger by a factor of 2, so the
coexistence field $s^*$ is commensurately smaller.)

It is instructive to generalize the mean-field analysis of the soft-FA model~\cite{Elmatad_PNAS_2010} to the East model.  In both soft-FA and soft-East models, the mean-field analysis gives the  correct low-temperature scaling for $s^*$.  However,
at this mean field level, one finds that the critical point scales as $s_\mathrm{c} = O(1)$ and $\epsilon_\mathrm{c} \sim \gamma$, as $\gamma\to0$.  
This analysis seems too simplistic, especially given that the mean-field analysis does not capture the hierarchical relaxation in the East model, and the mean-field predictions do not
seem consistent with the small values of $\epsilon_\mathrm{c}$ and $s_\mathrm{c}$  shown in Fig.~\ref{fig:phaseD}. 
A more appropriate estimate of $\epsilon_c$ can be obtained from dynamical scaling at equilibrium in the East model: scaling requires that
requires that $\tau_\mathrm{soft}/\tau_\mathrm{East}$ should remain constant as $\gamma$ is reduced.   
Assuming that the large deviations of $K$ have the same scaling as the equilibrium fluctuations, one therefore expects 
\begin{equation}
\epsilon_\mathrm{c} \approx \exp\left[ -\frac{(\log \gamma)^2}{2\ln 2}\right]
\label{equ:eps_c}
\end{equation}
with $s_\mathrm{c} \sim \gamma\epsilon_\mathrm{c}$ as constrained by the symmetry condition~(\ref{eqn:sym}). 
These predictions indicate that the existence of a space-time phase transition requires that the probability of constraint violation must remain 
extremely small at low temperature, at least in this model.

%%%%%%%%%%%%%%%%%%%%%%%%%%%%%%%%%%%%%%%%%%%%%%%%%%%%%%%%%%%%%
% Implications of These Results for the Glass Transition
%%%%%%%%%%%%%%%%%%%%%%%%%%%%%%%%%%%%%%%%%%%%%%%%%%%%%%%%%%%%%

\section{Implications of These Results for the Glass Transition}

\begin{figure}
\includegraphics[width=4cm]{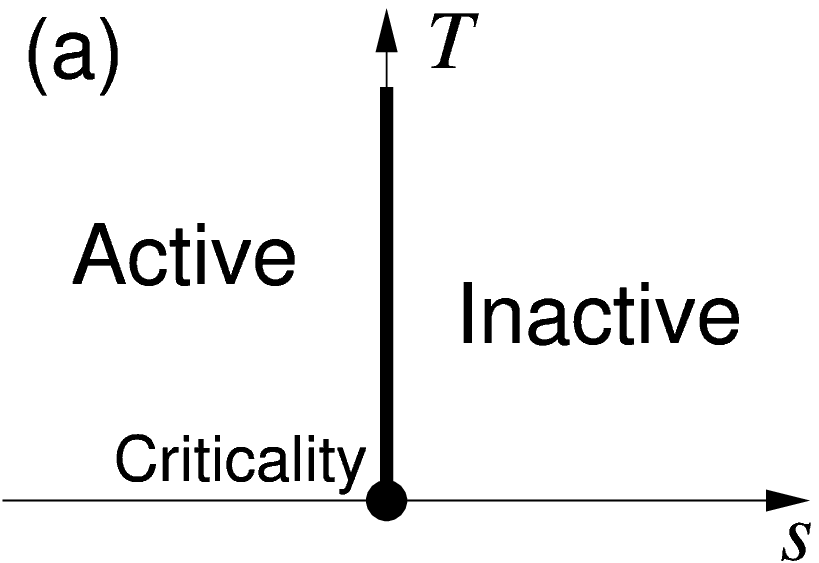}
\includegraphics[width=4cm]{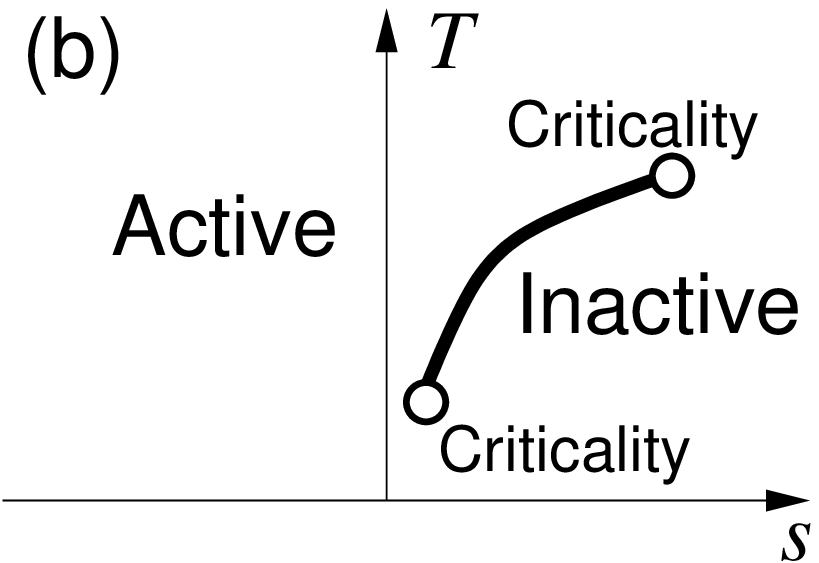}
\caption{Schematic phase diagrams in the $(s,T)$ plane.  (a)~The ``hard'' East model ($U\to\infty$, varying $T/J$ and $s$). 
The system is in the inactive phase for all $s>0$ and in the active phase for $s\leq0$. 
The equilibrium dynamics are characterized by diverging length and time scales as $T\to0$.
(b)~The softened East model with fixed $U/J>1$, varying $T/J$ and $s$.  For an intermediate range of temperature, 
there is a first order phase transition at $s>0$.  This is also the temperature range in which facilitation effects are strong.
Facilitation effects are weak both for large $T$ (above the onset temperature) and for small $T$ (below $T_\mathrm{x}$): there 
is no phase transition in these weak-facilitation regimes.
} \label{fig:phase}
\end{figure}

While there is no direct experimental realization of the $s$-field, we believe the results presented here have implications to real glassformers.  Again, we are motivated via analogy to the ferromagnetic system, as discussed in~\cite{Jack2006}.  Imagine a ferromagnet~\cite{IMSM} at temperature $T<T_\mathrm{c}$ and in a magnetic field $h > 0$.  The dominant phase is ordered and aligns with the field $h$.  However, there remain fluctuations which are aligned opposite to the preferred phase.  The size of these fluctuations and their shape are controlled by the free energy difference and surface tension between the two phases~\cite{IMSM}.  Near $T_\mathrm{c}$, the character of these fluctuations is influenced by the nearby critical point and the fluctuations become larger and their shape becomes less well defined. 

For `hard' KCMs (with $\epsilon = 0$), the coexistence between the active and inactive phases occurs exactly at $s^*=0$~\cite{Garrahan_PRL_2007}.  On the other hand, for the softened models presented here and in Ref. \cite{Elmatad_PNAS_2010}, the coexistence moves towards $s^* > 0 $.  The equilibrium dynamics of the system corresponds to the active phase which dominates the system.
However, for equilibrium dynamics close to coexistence, we still expect that this dominant phase may support relatively large fluctuations of the minority (inactive) phase, depending on the size of the free energy difference and the space time ``surface tension".  This was
the scenario envisaged in~\cite{Merolle_PNAS_Aug_2005,Jack2006}: domains of the inactive phase form the slow correlated regions that form part of the dynamical heterogeneity
in the glassy systems.

In the soft East model, we estimate the free energy difference between active (equilibrium) and inactive phases to be $\Delta \mu \approx s^* \langle k(s=0) \rangle \approx \gamma\epsilon$ per site, per unit time.  Both $\gamma = \ee^{-J/T}$ and $\epsilon = \ee^{-U/T}$ decrease on cooling, so this effect promotes fluctuations of the inactive phase, and enhanced dynamical heterogeneity.  However, we expect the low-temperature crossover at $T_\mathrm{x}$ to be relevant for the low-temperature behavior (recall Sec.~\ref{sec:model}).
Assuming that (\ref{equ:eps_c}) holds, then $\epsilon_\mathrm{c}$ tends to zero faster than $\epsilon \sim \ee^{-U/T}$ on cooling, with the result that $\epsilon > \epsilon_\mathrm{c}$ at low temperatures, and there is no space-time phase transition for $T\lesssim T_\mathrm{x}$

To summarize, facilitated dynamics dominate the system for $T_\mathrm{x} < T < T_\mathrm{o}$, and this is the regime in which space-time phase transitions are observed.  Outside this regime, the dynamics of the system have weaker correlations in space and time, and the response to the field $s$ is smooth, not singular.  
The situation as a function of temperature $T$ and bias $s$ is sketched in Fig.~\ref{fig:phase}.  However, as discussed in Sec.~\ref{sec:model}, the critical point in the vicinity of $T_\mathrm{x}$ is probably hard to observe in atomistic models of glass-formers, since supercooling the system to such a low temperature is likely to be very difficult.  

We end with a few  comments on the space-time phase diagram we have proposed in Fig.~\ref{fig:phase}.  Firstly, while two critical points appear when the phase diagram is plotted in these variables, phase diagrams in the $(\epsilon,s)$ plane for any fixed $\gamma$ are always of the form shown in Fig.~\ref{fig:phaseD}(a).  Except for the point where $\epsilon=0$ and $\gamma\to0$ (the original East model as $T\to0$), these critical points are all of the same type.  We expect these to be in the universality class of the Ising model in $(1+1)$ dimensions, as in the soft-FA case~\cite{Elmatad_PNAS_2010}.  (Settling the universality class of these transitions requires an assumption that the microscopic left-right asymmetry of the East model is irrelevant on very long length scales near the critical point: we have not been able to demonstrate this explicitly but it seems plausible from our numerical results).  Finally, we note that passage from active to inactive states and vice versa may be achieved by varying $s$ and $T$ in Fig.~\ref{fig:phase}, without ever passing through any phase transition line.  Such a situation requires that the symmetry properties of the active and inactive phases must be the same, and forbids any spontaneous breaking of translational invariance in the inactive phase.  This is consistent with the behavior of the soft-FA model: in the analogy with atomistic glass-formers it requires that the inactive phase be a liquid, and not a crystal or glass phase where non-trivial density profiles persist indefinitely.

%%%%%%%%%%%%%%%%%%%%%%%%%%%%%%%%%%%%%%%%%%%%%%%%%%%%%%%%%%%%%%%%%%%%%%%%%%%%%%%%%%%%%%%%%%%%%%%%%%%%%%
% ACKNOWLEDGMENTS
%%%%%%%%%%%%%%%%%%%%%%%%%%%%%%%%%%%%%%%%%%%%%%%%%%%%%%%%%%%%%%%%%%%%%%%%%%%%%%%%%%%%%%%%%%%%%%%%%%%%%%

\begin{acknowledgments}
We warmly thank D. Chandler and J. P. Garrahan for guidance and helpful discussions.  We further thank A. S. Keys and G. D\"uring for helpful comments on the manuscript.  
Y. S. E. was funded by the NSF GRFP and DOD NDSEG program during the beginning of this project and by an NYU Faculty Fellowship during the completion.  RLJ thanks the Engineering and Physical Sciences Research Council (EPSRC) for support through grant EP/I003797/1.
\end{acknowledgments}

%%%%%%%%%%%%%%%%%%%%%%%%%%%%%%%%%%%%%%%%%%%%%%%%%%%%%%%%%%%%%%%%%%%%%%%%%%%%%%%%%%%%%%%%%%%%%%%%%%%%%%
% BIBLIOGRAPHY
%%%%%%%%%%%%%%%%%%%%%%%%%%%%%%%%%%%%%%%%%%%%%%%%%%%%%%%%%%%%%%%%%%%%%%%%%%%%%%%%%%%%%%%%%%%%%%%%%%%%%%

\end{document}